\documentstyle[epsf]{ioplppt} 

\newfont{\BlackBoardBold}{msbm10}
\def\BBB{\fam14 }
\def\Re{{\BBB R}}

\def\be{\begin{equation}}
\def\ee{\end{equation}}
\def\bea{\begin{eqnarray}}
\def\eea{\end{eqnarray}}

\begin{document}
\title{Evolution of the density contrast in inhomogeneous dust models}
\author{Filipe C. Mena $\natural$ $\flat$\ftnote{3}{E-mail:fcm@maths.qmw.ac.uk} 
and Reza Tavakol $\natural$\ftnote{4}{E-mail:reza@maths.qmw.ac.uk}}

\address{$\natural$\  Astronomy Unit,
                School of Mathematical Sciences,
                Queen Mary \& Westfield College,
                Mile End Road,
                London E1 4NS, UK}
\address{$\flat$\     Departamento de Matem\'{a}tica,
                      Universidade do Minho,
                      Campus de Gualtar,
                      4710 Braga,
                      Portugal}

\maketitle
\begin{abstract}
With the help of families of density contrast indicators,
we study the tendency of gravitational systems to become increasingly
lumpy with time. Depending
upon their domain of definition, these 
indicators could be
local or global. We make a comparative study
of these indicators in the context of inhomogeneous
cosmological models of Lemaitre--Tolman and Szekeres. In particular,
we look at the temporal asymptotic behaviour of these
indicators and ask under what conditions,
and for which class of models, they evolve
monotonically in time.

We find that for the case of ever-expanding models,
there is a larger class of indicators that grow monotonically with time, whereas
the corresponding class 
for the recollapsing models is more restricted. 
Nevertheless, in the absence of decaying modes,
indicators exist
which grow monotonically with time for both
ever-expanding and recollapsing models
simultaneously.
On the other hand, no such indicators may
found which grow monotonically if the decaying modes are allowed 
to exist.
We also find the conditions
for these indicators to be non-divergent
at the initial singularity in both
models.

Our results can be of potential relevance
for understanding structure formation in 
inhomogeneous settings and in debates
regarding gravitational entropy
and arrow of time. In particular,
the spatial dependence of turning
points in inhomogeneous cosmologies
may result in multiple density contrast arrows in 
recollapsing models over certain epochs. We also find that
different notions of asymptotic homogenisation
may be deduced, depending upon
the density contrast indicators used.
\end{abstract}
\section{Introduction}
An important feature of classical self-gravitating systems
is that in general they are not
in equilibrium states. This instability gives rise to
spontaneous creation of structure which is assumed
to lead to
{\it increasing} lumpiness as time increases.
 
This rise of structure with time
in self-gravitating systems seems to run counter
to the usual intuition in classical statistical
physics where increasing time is associated with the 
increase in microscopic disorder and hence to
macroscopic uniformity
in matter. The main reason for this difference
is thought to be due to the long range and
unshielded nature of the gravitational force \cite{Padmanabhan90}.
This dichotomy
has led to the question of possible existence 
of gravitational entropy
and its connection with the usual thermodynamic 
entropy (see for e.g. \cite{Penrose79,Book-Arrow-Time} and references
therein).
 
Whether a satisfactory notion of gravitational
entropy exists and whatever its nature may be,
it is of interest to find out the extent
to which the assumption regarding
the increase in
structuration with increasing time
holds in relativistic cosmological models. This would be of value
for a variety of reasons, including its potential relevance
for debates concerning structure formation in the Universe.
Furthermore, the presence or absence of indicators
that evolve monotonically in time
could inform the debates regarding gravitational
entropy and in particular whether, if it in fact exists,
it should be
monotonic in models which possess recollapsing phases
(see for e.g. \cite{Hawking85,Page,Hawkingetal93}).

Since indicators that best codify such structuration
are not known a priori, rather than
focusing on a single indicator, we shall here consider families
of spatially covariant indicators which measure the density contrast
and which include as special cases
indicators 
put forward by
Bonnor \cite{Bonnor86} and Tavakol \& Ellis \cite{Tavakol-Ellis}.
We shall also consider for completeness some
indicators previously used in the literature, 
including that given by Szafron \& Wainwright \cite{Szafron-Wainwright}
and the non-covariant indicators of
Silk \cite{Silk} and
Bonnor \cite{Bonnor74},
even though the latter two may, in view of their lack of covariance, be viewed as 
suspect from a physical point of view.
\\
 
We shall employ inhomogeneous cosmological models
of Lemaitre--Tolman (LT) \cite{Lemaitre,Tolman} 
and Szekeres \cite{Szekeres}
in order to make a
comparative study of these indicators in the 
general relativistic inhomogeneous settings.
Since these models involve arbitrary\footnote{These functions
are not in fact totally arbitrary as they need
to satisfy certain constraints which we shall
discuss below.}
functions,
the integrals involved in the definitions of
our indicators cannot be performed in general.
As a result, we look at the asymptotic behaviour
of both ever-expanding and
recollapsing families of models as well as 
their behavious near the origin,
and in particular look for
conditions under which the
asymptotic evolution of these indicators
is monotonic with time.
Clearly these can only give
necessary conditions for the all-time
monotonic evolution
of these indicators. To partially extend these results to
all times we also calculate these indicators
for a number of concrete models given in the literature.
The crucial point is that our asymptotic results 
bring out some general points that seem
to be supported by our all-time study of 
the concrete models.
\\
 
The organisation of the paper is as follows. In section 2
we introduce and motivate families of density contrast indicators.
Sections 3--6 contain the calculations of these
measures for LT and Szekeres models respectively. 
In section 7 we consider the behaviour of the 
dimensionless analogues of these indicators. In section 8 we 
study the behaviour of these indicators
near the initial singularity. Section
9 gives 
a discussion of some of the consequences of
our results and
finally section 10 contains our conclusions.
\\

Throughout we use units in which $c=G=1$ and lower case
latin indices take values 0 to 3.
\section{Density contrast indicators}
\label{DC-Indicators}
Intuitively one would expect the rise of structuration
in cosmology to be related to the
coarse grained spatial variance of various
physical scalars.
An important scalar in cosmology
is the energy density $\rho$ (or in a multi-component
version of this, densities $\rho^{(i)}$
corresponding to the different components of the content
of the Universe). Here for simplicity
we confine ourselves to the one component case and
consider the rotation-free dust setting. We
can then
introduce a global spatial variability index defined
as
\be
{\Large\int} \frac{\left |\rho -
{\rho}_0 \right |}{
\rho_0} dV
\ee
where $\rho_0$ is the mean density defined appropriately
and $dV$ is the comoving volume element.
One can make this notion spatially covariant by, for example, expressing it
in terms of the fractional density gradient
introduced by Ellis $\&$ Bruni \cite{Ellis-Bruni},
\be
\chi_{a}= \frac{h_a^b}{\rho}\frac{\partial\rho}{\partial x^b}
\label{def}
\ee
where $h_{ab} = g_{ab} + u_a u_b $ projects orthogonal to the unit
4-velocity $u^a$, which we shall throughout assume to be 
uniquely defined.
A
related covariant spatial variability index can then
be defined thus
\begin{equation}
\int_\Sigma \left |\chi_{a} \right | dV
\label{index}
\end{equation}
where the integration is over a 3-surface $\Sigma$ or part thereof. 
\\
 
Now it is not a priori clear what are the indicators
that best codify such structuration and in particular
what their monotonic properties may be. As a result,
instead of 
concentrating on a single indicator, we shall 
introduce a two parameter
family of possible covariant indicators, which we refer to
as {\it density contrast indicators}, $S_{IK}$,
in the form
\be
\label{index-rewrite}
S_{IK}=
\int_\Sigma  \left | \frac{h^{ab}}{\rho^I}
\frac{\partial \rho}{\partial x^a}
\frac{\partial \rho}{\partial x^b}\right |^{K} dV, ~~~~I\in\Re,~K \in \Re 
\setminus \{0\}.
\ee
An important feature of this family is that it may be treated as
local or global depending upon the
size of $\Sigma$. It also includes as special
cases the indicator given by Tavakol \& Ellis \cite{Tavakol-Ellis},
for which $I=2$ and $K=1/2$, and the pointwise 
indicator previously given 
by Bonnor \cite{Bonnor86}
\be
B1 =\frac{h^{ab}}{\rho^2}
\frac{\partial \rho}{\partial x^a}
\frac{\partial \rho}{\partial x^b}.
\ee
In the cosmological context we might expect it
to be more appropriate to normalise these indicators
with the comoving volume. We therefore define
the corresponding density contrast indicators per unit volume,
$S_{IKV}$, by
\be
\label{comoving-ind}
S_{IKV}=\frac{S_{IK}}{V}
\ee
where $ V = \int_\Sigma dV$ is the comoving volume.
We shall also consider dimensionless analogues
of these indicators in section 7.
 
Indicators (\ref{index-rewrite}) and (\ref{comoving-ind})
are of potential interest for a number of reasons,
including their operational definability and hence 
their potential relevance to
observations regarding the evolution of structure
in the Universe
and their possible connection to the question of gravitational entropy.
\\

For completeness we shall also consider
the spatially covariant indicator introduced by
Szafron \& Wainwright  \cite{Szafron-Wainwright}
\be
SW = -\frac{1}{\dot\rho}\sqrt{h^{ab}
\frac{\partial \rho}{\partial x^a}
\frac{\partial \rho}{\partial x^b}}
\ee
as well as the non-covariant indicators given for
LT models by
Bonnor \cite{Bonnor74}
\be
B2  = \frac{1}{\rho}\frac{\partial \rho}{\partial r}
\ee
and 
Silk \cite{Silk}
\be
SL= \frac{r}{\rho}\frac{\partial \rho}{\partial r}
\ee
where $r$ in these expressions is the $r$ 
coordinate of the LT models introduced in the next
section. We note that some of these latter indicators 
have been used as measures of homogeneity in
the past \cite{Bonnor74,Szafron-Wainwright}, a question we shall return to 
section 9.
\\

In the following, in analogy with the notion
of {\it cosmological arrow} which points in the 
direction of the dynamical evolution of the Universe,
we shall employ the notion of
{\it density contrast arrow} which is in the
direction of the evolution of the
density contrast indicator employed.
 
The aim here is to test these families of
indicators in the context
of
LT and Szekeres models
in order to determine the subset of these indicators (and models) 
for which the asymptotic evolution and the evolution near the
origin is 
monotonically increasing with time, i.e there is
a unique density contrast arrow which points in the direction
of increasing time.
\section{Lemaitre--Tolman models in brief}
The Lemaitre--Tolman models \cite{Lemaitre, Tolman, Bondi} (see also                   
\cite{Krasinski})
are given by
\be
ds^2 = -dt^2 + \frac{{R^{'}}^2}{1+f} dr^2 +R^2 (d\theta^2 + \sin^2
\theta
d\phi^2)
\label{tolman}
\ee
where $r, \theta, \phi$ are the comoving coordinates, $R=R (r,t)$ and $f=f(r)$ are arbitrary $C^2$ real functions
such
that $f > -1$ and $R (r,t)\ge 0$. In this section
a dot and a prime 
denote $\partial / \partial t$ and $\partial / \partial r$ respectively.
The evolution of these models is then given by
\be
\label{tol-eq}
{\dot R}^2 = \frac{F}{R} +f
\ee
where $F=F(r)$ is another $C^2$ arbitrary real function, assumed
to be positive in order to ensure the positivity
of the gravitational mass.
Equation (\ref{tol-eq})
can be solved 
for different values of $f$
in the following
parametric forms:
\\

\noindent {For $f<0$}:

\bea                                                               
\label{elliptic}
&R =  \frac{F}{2(-f)}(1-\cos\eta) \nonumber  \\  
&(\eta-\sin\eta)  =  \frac{2(-f)^\frac{3}{2}}{F}(t-a)
\eea
where $0< \eta < 2\pi$ and $a$ is a third arbitrary real function of $r$.
\\

\noindent {For $f>0$}:
\bea
\label{hyperbolic}
&R = \frac{F}{2f}(\cosh\eta-1), \nonumber  \\ 
&(\sinh\eta-\eta) = \frac{2f^\frac{3}{2}}{F}(t-a)
\eea
where $\eta > 0$.
\\

\noindent {For $f=0$}:

\be
\label{parabolic}
R=\left(\frac{9F}{4}\right)^\frac{1}{3}(t-a)^\frac{2}{3}.
\ee
The solutions corresponding to $f>0$, $f=0$ and $f<0$ are referred  
to as hyperbolic,
parabolic and elliptic, respectively. In the elliptic case,
there is a recollapse to a second singularity, while the
other two classes of models are ever-expanding.
In all three cases the matter density can be written as
\be
\label{density}
\rho(r,t) = \frac{F^{'}}{8\pi R^{'} R^2}.
\ee
Now the fact that $\rho$ must be non-divergent (except on 
initial and final singularities) and positive everywhere imposes 
restrictions 
on the arbitrary functions \cite {Hellaby85}, with the positivity
of the density 
implying that $R'$ and $F'$ have the same sign.
\section{Evolution of the density contrast in Lemaitre--Tolman models}
For these models the indicators (\ref{index-rewrite}) can be written
as 
\be
\label{entropy}
S_{IK} = 4\pi \int \left |\frac{1+f}{R'^2}\frac{1}{\rho^I} \left (\frac{\partial
\rho}{\partial r} \right)^2 \right |^K \frac{R^2 |R'|}{\sqrt{1+f}}  dr
\ee
with the
time derivative 
\bea
\label{rate}
\dot{S}_{IK}  =  4\pi &\int & (1+f)^{K-\frac{1}{2}}
\left [ \frac{\partial}{\partial t} \left(R^2|R'|^{1-2K}\right) \frac{1}{\rho^I}
\left(\frac{\partial\rho}{\partial r}\right) \right. \nonumber  \\ 
& + & \left. R^2|R'|^{1-2K} \frac{\partial}{\partial t}
\left(\frac{1}{\rho^I} \left(\frac{\partial \rho}{\partial r}
\right)^2\right)  \right] dr.
\eea
In the following we consider 
different classes of LT models given by different
types of $f$. 
Clearly in general
$\rho$ in such models depends on the  
functions $f$ and $F$ and as a result the integrals            
arising in $S_{IK}$ and $S_{IKV}$ cannot be performed in general. 
We shall therefore look at the 
asymptotic behaviour of these indicators
and in some special cases we study the behaviour of the indicators 
for all times. 

We note that in some studies concerning the 
question of gravitational entropy the function $a$ has been taken
to be a constant or zero (see for example Bonnor \cite{Bonnor85}),
in order to avoid the presence of white holes.
Here, in line with these studies, we shall also
take $a$ to be a constant in all sections below,
apart from section (9.2) where the shortcomings 
associated with taking 
$a = a(r)$ will not affect our 
results.
\subsection{Parabolic LT models}
Models of this type
with  $a=$ constant, reduce to the Einstein--de Sitter model \cite {Bonnor74}
for which  the density depends only on $t$
giving 
$S_{IK}=0=S_{IKV}$ for all time.
\subsection{Hyperbolic LT models}
\label{LT-Hyperbolic}
For large $\eta$ we have 
\be
\label{r6}
R\approx f^{\frac{1}{2}}(t-a)
\ee
which gives
\bea
\rho &\approx &  \frac{F'}{4\pi f'f^{\frac{1}{2}}(t-a)^{3}}\\ 
\dot{\rho} &\approx &  \frac{-3 F'}{4\pi 
f'f^{\frac{1}{2}}(t-a)^4}
\eea
where for positivity of $\rho$, $F'$ must have the same
sign
as $f'$.
This then gives 
\bea
S_{IK} & \approx & 4\pi
\int \alpha_1 (t-a)^{3IK-8K+3}dr \\
S_{IKV} & \approx & 2\frac{\int \alpha_1 (t-a)^{3IK-8K} dr}{\int |f'|f^{\frac{1}{2}}(t-a)^3 dr}
\eea
where
$\alpha_1>0$ is purely a function of $r$.

The dominant asymptotic temporal behaviour
of the density contrast
indicators described in section 2 are calculated and
summarised in the column 2 of 
Table (\ref{Indicators-Tolman}) and the conditions for $S_{IK}$ and $S_{IKV}$
to be monotonically increasing
can then be readily calculated and are summarised in
Table (\ref{DC-Tolman}).
\subsection{Elliptic LT models}
In the limit $\eta\to\ 2\pi$, $R$, $\rho$ and 
$\dot\rho$ can be written as
\bea
R & \approx  & \phi_{1}\phi_{2}^{\frac{2}{3}} \\
\rho & \approx & \frac{3F'}{16\pi\phi_1 \phi_2'\phi_2}\\ 
\dot{\rho} &\approx & \frac{27F'(-f)^{\frac{3}{2}}}{\pi\phi_1 F}
\left(\frac{F'}{F}-\frac{3}{2}\frac{f'}{f}\right)
\frac{\phi_2+\frac{(-f)^{\frac{3}{2}}}{F}t}{\phi_2'^2\phi_2^2}
\eea
where  
$\phi_{2}(r,t)=12\left [\frac{(-f)^{\frac{3}{2}}}{F}(t-a)-\pi\right]$, 
$\phi_{2}'(r,t)=-12\frac{(-f)^{\frac{3}{2}}}{F}\left[
\left(\frac{F'}{F}-\frac{3}{2}\frac{f'}{f}\right)t\right]$ and \\ 
$\phi_{1}(r)=\frac{F}{4(-f)}$.
Now $\phi_{2}$ satisfies 
$\phi_{2}<0$
and $\dot{\phi_{2}}>0$ and $\phi_2'$ and $F'$ must have opposite 
signs to ensure the positivity of $\rho$. The indicators are
then given by
\bea
S_{IK} &\approx  &4\pi \int \alpha_2 |\phi_2|^{IK-\frac{10}{3}K+1} dr \\
\dot{S}_{IK} & \approx & 4\pi \int
\left(IK-\frac{10}{3}K+1\right)\alpha_2 \phi_2'
|\phi_2|^{IK-\frac{10}{3}K} \nonumber \\
&+&\dot{\alpha_2}' |\phi_2|^{IK-\frac{10}{3}K+1} dr
\eea
where $\alpha_2(r,t)=\frac{2}{3}|\phi_2'| \phi_1^3 \left|\left(\frac{3}{2}\right
)^4\frac{(1+f)F'^2}{\phi_1^8 \phi_2'^2}\left(\frac{2\phi_1^3 \phi_2'}{3F'}\right)^I\right|^K >0$.
Similarly the density contrast per unit volume can be calculated to be
\be
S_{IKV}=\frac{3}{2}\frac{\int \alpha_2 |\phi_2|^{IK-\frac{10}{3}K+1}dr}{\int
\phi_1^3 |\phi_2| dr}.
\ee
Now with the choice of $\left(\frac{F'}{F}-\frac{3}{2}\frac{f'}{f}\right) =0$, 
the model becomes homogeneous with $S_{IK} =0 = S_{IKV}$. 
When $\left(\frac{F'}{F}-\frac{3}{2}\frac{f'}{f}\right)\neq0$,
we have 
$\frac{\partial |\phi'_2|}{\partial t}>0$
and $\frac{\partial |\phi_2|}{\partial t}<0$ and 
our results are summarised
in Tables (\ref{Indicators-Tolman}) and (\ref{DC-Tolman}).
We note that the asymptotic behaviour of
$S_{IKV}$ cannot be deduced in general
and therefore the corresponding entries in these tables
were derived using pointwise
versions of the indicators.

\begin{table}[!htb]
\begin{center}
\begin{tabular}{cll}
\hline
Indicators & $f>0$ & $f<0$ \\
\hline
\\
$S_{IK}$ &  $(t-a)^{3IK-8K+3}$ &
$\phi_2^{IK-\frac{10}{3}K+1}$\\
\\
$S_{IKV}$ & $(t-a)^{3IK-8K}$
 & $\phi_2^{IK- \frac{10}{3} K}$\\
\\
$B1$&  $(t-a)^{-2}$
& $\phi_2^{-1/3}$\\
\\
$B2$&
 $const.$ & $\phi_2^{-1}$\\
\\
SL & 
$const.$ & $\phi_2^{-1}$\\
\\
SW & $const.$
& $\phi_2^{1/3}$\\
\\
\hline
\end{tabular}
\caption[Indicators-Tolman]{\label{Indicators-Tolman}Asymptotic evolution of
density contrast indicators given in section (\ref{DC-Indicators}), for 
hyperbolic and elliptic LT models. The constants in the second
column are different for each $r$.} 
\end{center}
\end{table}

\begin{table}[!htb]
\begin{center}
\begin{tabular}{ccc}
\hline
~Models~~~~~ & $~~~~~S_{IK}~~~~~$ & $~~~~~S_{IKV}~~~~~$ \\
\hline
\\
$f >0 $ & $I>\frac{8}{3}-\frac{1}{K}$ & $I>\frac{8}{3}$\\
\\
$f <0 $ & $I<\frac{10}{3}-\frac{1}{K}$ & $I<\frac{10}{3}$\\
\\
\hline
\end{tabular}
\caption[DC-Tolman]{\label{DC-Tolman} Constraints on $I$ and $K$
in order to ensure 
$\dot{S}_{IK}>0; \dot{S}_{IKV}>0$ asymptotically in 
hyperbolic and elliptic LT models.}
\end{center}
\end{table}
\vskip .2in
To summarise, the results of this section indicate
that for both ever-expanding ($f>0$) and
recollapsing ($f<0$) LT models,
$I$ and $K$ can always be chosen such that $S_{IK}$ and $S_{IKV}$
both grow asymptotically.
However, there are special cases of interest,
such as $I=2$, for which no such
intervals can be found.
\subsection{Special LT models}
\label{Special}
 So far we have studied the behaviour of the $S_{IK}$ and $S_{IKV}$
asymptotically. To partly extend these results 
to all times,
we shall in this section consider some concrete examples 
of LT models which have been considered in the literature.
\\

\noindent{\em Parabolic examples}:
\\\\
Models
with $a=0$ (see e.g. those
in \cite{Bonnor74} and \cite{Maartens})
are 
homogeneous with trivial behaviour for the indicators.
\\

\noindent{\em Hyperbolic examples}:
\\\\
We considered a number of examples of this type
given by Gibbs \cite{Gibbs}, Humphreys et al. \cite{Maartens}, 
and Ribeiro \cite{Ribeiro93},
the details of which are summarised 
in Table (\ref{tablehyperbolic}).
We found that for all these models 
there exist
ranges of $I$ and $K$ such that
indicators $S_{IK}$ increase monotonically for all time.
In particular, we found that the condition
for all time monotonicity with $K=1/2$ and $K=1$ are
given by $I \in \left[1, +\infty\right[$ and
$I \in \left[2,+\infty \right[$ respectively.
\\

\begin{table}[!htb]
\begin{center}
\begin{tabular}{clll}
\hline
References & $F(r)$ & $f(r)$ \\
\hline
\\
Humphreys et al. & $F=\frac{1}{2}r^4$ & $f=r^2$  \\
\\
Humphreys et al. & $F=\frac{1}{2}r^3$ & $f=r^3$ \\
\\
Gibbs  & $F=F_0 \tanh r^3$ & $f=f_0 \sinh^2 r$ &\\
\\
Ribeiro & $F=F_0 r^p$  & $f=\sinh^2 r $ \\
\\
\hline
\end{tabular}
\caption[tablehyperbolic]{\label{tablehyperbolic}Examples of 
hyperbolic LT models, where
$F_0, p$ and $f_0$ are
positive constants.}
\end{center}
\end{table}

\noindent{\em Elliptic examples:}
\\\\
We considered examples of this type,
given by Bonnor \cite{Bonnor85c} 
and Hellaby \& Lake \cite{Hellaby85},
details of which are summarised in Table (\ref{tableelliptic}).
Again we found that for all these models
there exist
ranges of $I$ and $K$ such that
indicators $S_{IK}$ are monotonic for all time.
In particular, we found that the condition
for all time monotonicity with $K=1/2$ and $K=1$ are
given by $I \in \left ]0, 1 \right ]$ and
$I \in \left]0, 2 \right]$ respectively.
For values of $K=1/2$ and $I$ outside the range $I< 4/3$, e.g. $I=2$,
we find that $S_{I\frac{1}{2}}$ increases in the
expanding phase while decreasing after a certain time (which depends on
$r$)
in the contracting phase and tending to zero
as the second singularity
approaches.
\begin{table}[!htb]
\begin{center}
\begin{tabular}{clll}
\hline
References & $F(r)$ &  $f(r)$ \\
\hline
\\
Bonnor & $F=F_0 r^3$ & $f=-f_0\frac{r^2}{1+r^2}$ \\
\\
Hellaby \& Lake & $F=F_0 \frac{r^m}{1+r^n}$ &  $f=-f_0 \frac
{r^n}{1+r^n}$ \\
\\
\hline
\end{tabular}
\caption[tableelliptic]{\label{tableelliptic}Examples of elliptic
LT models, where 
$F_0$ and $f_0 \ne 1$ are positive real constants and $m$, $n$ are integers such
 that $m>n$.}
\end{center}
\end{table}
\\

To summarise, we have found that for all these concrete examples
there exists values of $I$ and $K$ such that
indicators $S_{IK}$ increase monotonically for all time.
However, the allowed intervals of $I$ and $K$ are in general narrowed
relative to those obtained by
the asymptotic considerations.
Finally the indicators $S_{IKV}$ in all these models 
lead to non elementary
integrals.
\section{Szekeres models in brief}
\label{Szekeres}
The Szekeres metric is given by \cite{Szekeres,Goode-Wainwright}
\be
\label{Metric-Szekeres}
ds^2=-dt^2+R^2e^{2\nu}(dx^2+dy^2)+R^2H^2W^2dz^2
\ee
where $W=W(z)$ and $\nu=\nu (x,y,z)$ are functions to
be 
specified within each Szekeres class and $R=R(z,t)$
obeys the evolution equation
\be
\label{evolution}
\dot{R}^2=-k+2\frac{M}{R},
\ee
where $M=M(z)$ is a positive arbitrary function for the 
class I models and a positive constant for class II models,
defined below.
The function $H$ is given by
\be
H=A-\beta _{+}f_{+}-\beta _{-}f_{-}
\ee
where functions $A =A(x,y,z)$, 
$H$, $R$ and $W$ are assumed to be positive;
$\beta _{+}$ and $\beta _{-}$ are functions of $z$
and $f_{+}$ and $f_{-}$ are functions of $z$ and $t$, corresponding to the
growing
and decaying modes of the solutions $X$ of the equation
\be
\label{perturbation}
\ddot{X}+\frac{2\dot{R}\dot{X}}{R}-\frac{3MX}{R^3}=0.
\ee
The density for these models is given by
\be
\label{Density-Szekeres}
\rho(x,y,z,t)=\frac{3MA}{4\pi R^3H}.
\ee
The solutions to (\ref{evolution}) are given by 
\bea
\label{solevolution}
&R=M\frac{dh(\eta)}{d\eta} \nonumber \\
&t-T(z)=Mh(\eta)
\eea
where
\be
h(\eta)=\left\{\begin{array}{lll}
                          \eta-\sin\eta & (k=+1) &0<\eta<2\pi \\
                          \sinh\eta-\eta & (k=-1) & 0<\eta \\
                          \frac{1}{6}\eta^3 & (k=0) & 0<\eta,
                     \end{array} 
               \right. 
\ee
which in turn allows the solutions to (\ref{perturbation}) to be written as
\bea
\label{fplus}
f_{+}(\eta)&=&\left\{\begin{array}{lll}
                   6MR^{-1}(1-\frac{1}{2}\eta\cot\frac{1}{2}\eta)-1 &
k=+1  \\
                   6MR^{-1}(1-\frac{1}{2}\eta\coth\frac{1}{2}\eta)+1  &
k=-1\\
                          \frac{1}{10}\eta^2 & k=0 
                     \end{array}
               \right. \\
f_{-}(\eta)&=&\left\{\begin{array}{lll}
                          6MR^{-1}\cot\frac{1}{2}\eta & k=1\\
                          6MR^{-1}\coth\frac{1}{2}\eta & k=-1  \\
                          24\eta^{-3} & k=0.
                     \end{array} 
               \right.
\eea
The Szekeres models are divided into two classes, depending upon 
whether $\frac{\partial (Re^\nu)}{\partial z}\ne 0$ or 
$\frac{\partial (Re^\nu)}{\partial z}=0$.
The functions in the metric take different forms for each class 
and for completeness are
summarised in the Appendix.
\section{Evolution of the density contrast in Szekeres models}
In these models the density contrast indicators are given by
\bea
\label{Entropy-Szekeres}
 S_{IK}=
&\int_\Sigma& \left |
\frac{1}{e^{2\nu}R^2\rho^I}\left(\frac{\partial \rho}{\partial
x}\right)^2+\frac{1}{e^{2\nu}R^2\rho^I}
\left(\frac{\partial \rho}{\partial
y}\right)^2 \right. \nonumber \\
&+& \left. \frac{1}{H^2W^2R^2\rho^I}
\left(\frac{\partial \rho}{\partial z}\right)^2
\right |^KR^3e^{2\nu}HWdxdydz
\eea
We note that the singularities
in these models are given not 
only by $R=0$ but also by $H=0$, which define the so called
shell crossing singularities.
For example, choosing $\beta_+>0$ 
eventually results in 
a shell crossing  singularity  at a finite time given by
$A=\beta_+f_+$ (see \cite{Goode-Wainwright} for 
a detailed discussion). Here in order to avoid shell
crossing singularities, we either assume  $\beta_+<0$ in all cases, or
alternatively $\beta_+ >0$ and
$\beta_+(z)<A(x,y,z)$, in the $k=-1$ case. 
\\

We recall that for class II models, $T=$ constant.
In the following we shall, in line with the work of
Bonnor \cite{Bonnor86}, also make this assumption in
the case of
class I models (implying $\beta_-=0$) in order to make the
initial singularity simultaneous and hence
avoid
white holes.
\\

We consider the 
two
Szekeres classes in turn and in each case we study the 
three subclasses referred to as hyperbolic, parabolic and elliptic,
corresponding to the Gaussian curvatures $k=-1,0,+1$ 
respectively.
\subsection{Evolution of the density contrast in class I models}
Assuming $\beta_+ =0$ in this class makes 
$\rho=\rho(t)$\footnote{In fact, the Szekeres models, with $T=$ constant, reduce 
to FLRW iff $\beta_+=\beta_-=0$ \cite{Goode-Wainwright}.},
 which implies $S_{IK}=0=S_{IKV}$ for all time.
We shall therefore assume $\beta_+$ to be non-zero
and possibly $z$ dependent. The contribution of $\beta_-$ would be important 
 near the initial and final singularities but is irrelevant for the asymptotic
behaviour of parabolic and hyperbolic models.
\\

\noindent {\em Parabolic class I Szekeres models}:
\\

For the models of this type
with  $T=$ constant, the density depends only on $t$
which trivially gives
$S_{IK}=0=S_{IKV}$ for all time.
\\

\noindent {\em Hyperbolic class I Szekeres models}:
\label{Subsection-SzekeresI-Hyperbolic}
\\

For large $\eta$ we have
\bea
R & \approx & t-T \\
f_+ & \approx & \frac{-6\eta}{e^{\eta}}+1
\eea
resulting 
in
\bea
\rho  &\approx  &\frac{3MA}{4\pi (A-\beta_+)(t-T)^3}\\ 
\dot{\rho} &\approx  &\frac{-9MA}
{4\pi (A-\beta_+)(t-T)^4}  
\eea
which have the same $t$ dependence as the 
corresponding LT models.
Using (\ref{Entropy-Szekeres}) we obtain
\bea
S_{IK} & \approx &  \int \alpha_4 (t-T)^{3KI-8K+3} dxdydz\\
S_{IKV} & \approx & \frac{\int \alpha_4 (t-T)^{3KI-8K+3} dxdydz}
{\int e^{2\nu} W (A-\beta_+) (t-T)^{3} dxdydz}
\eea
where $\alpha_4$ is a positive function of $x,y,z$.
\\

The dominant asymptotic temporal behaviour
of the density contrast
indicators 
for these models and the conditions for their
monotonic behaviour 
with time are summarised in Tables (\ref{Indicators-Szekeres}) and
(\ref{DC-Szekeres}).
\\

\noindent {\em Elliptic class I Szekeres models:}
\\\\
Using (\ref{Density-Szekeres}), together 
with the following approximations
\bea
R & \approx & \frac{6^{\frac{2}{3}}M\psi_2^{\frac{2}{3}} }{2}\\
f_+ & \approx & \frac{-6}{\psi_2}
\eea
where $\psi_2=\left(\frac{t-T}{M}-2\pi\right)$,
gives
\bea
\label{rho-zek+1}
\rho  &\approx  &\frac{6A}{\pi M^2\beta_+\psi_2}\\ 
\dot{\rho} &\approx  &\frac{-6A}
{\pi M^3 \beta_+\psi_2^2}
\eea
which has again the same temporal dependence as for the corresponding
closed LT models.
Using (\ref{Entropy-Szekeres}) we obtain
\bea
S_{IK} & \approx & \int \alpha_5 \psi_2^{IK-\frac{10}{3}K+1} dxdydz\\
S_{IKV} & \approx & \frac{2}{9}\frac{\int \alpha_5 \psi_2^{IK-\frac{10}{3}K+1} dxdydz}
{\int  M e^{2\nu} W \beta_+ \psi_2 dxdydz},
\eea
where\\ $\alpha_5 
(x,y,z,t)=\frac{9}{2}MWe^{2\nu}\pi\beta_+
\left|\left(\frac{96^2 48^{-I}6^{-\frac{4}{3}}A^{-I}e^{-2\nu}}
{M^{\frac{14}{3}-2I}
(\pi\beta_+)^{2-I}}\right)
\left(
\left(\frac{\partial A}{\partial x}\right)^2+
(\frac{\partial A}{\partial 
y})^2+\frac{e^{2\nu}\psi_2^{'2}}{W^2(\pi\beta_+)^2}\right)
\right|^K$, and as $\eta\to 2\pi$, $t\to 2\pi M+T$ and $\psi_2' \to
 -2\pi M'$.
\\

Our results are
summarised in Tables (\ref{Indicators-Szekeres}) and
(\ref{DC-Szekeres}).
We note that the asymptotic behaviour of
$S_{IKV}$ cannot be deduced in general
for these models and therefore the corresponding entries in these tables
were derived using pointwise
versions of the indicators.
\subsection{Evolution of the density contrast in class II models}
Recall that in this class  $T$
and $M$ are constants in all cases. Again if $\beta_+=\beta_-=0$ we recover the 
homogeneous models giving $S_{IK}=0=S_{IKV}$. In what follows we shall therefore consider
the general cases with $\beta_+=\beta_+(z)$ and $\beta_-=\beta_-(z)$.
\\

\noindent {\em Parabolic class II Szekeres models:}
\\\\
The asymptotic evolution of $\rho$ in this case is 
given by 
\bea
\rho & \approx & \frac{5A}{3\pi(-\beta_+)(t-T)^{\frac{8}{3}}} \\
\dot{\rho} & \approx & \frac{40A}
{9\pi \beta_+(t-T)^{\frac{11}{3}}}.
\eea
Using (\ref{Entropy-Szekeres}) we obtain
\bea
S_{IK}  & \approx & \int \alpha_6(t-T)^{\frac{8}{3}KI-
\frac{20}{3}K+\frac{8}{3}} dxdydz\\
S_{IKV}  & \approx &  \frac{2}{9}\frac{\int \alpha_6 (t-T)^{\frac{8}{3}KI-
\frac{20}{3}K+\frac{8}{3}} dxdydz}
{\int e^{2\nu} |\beta_+|  M W (t-T)^{\frac{8}{3}}dxdydz}
\eea
where $\alpha_6$ is a positive function of $x,y$ and $z$ only.
Our results are shown in 
Tables (\ref{Indicators-Szekeres}) and
(\ref{DC-Szekeres}).
\\

\noindent {\em Hyperbolic class II Szekeres models:}
\\\\
Here we use the approximations for $R, H, f_+$ and $f_-$ already shown for class I.
The asymptotic evolution of $\rho$ and $\partial\rho / \partial t$ are 
in this case given by 
\bea
\rho & \approx & \frac{6MA}{\pi(A-\beta_+)(t-T)^3} \\
\dot{\rho} & \approx & \frac{-18MA}
{\pi (A-\beta_+)(t-T)^4}
\eea
Using (\ref{Entropy-Szekeres}) we obtain
\bea
S_{IK}  & \approx & \int \alpha_7 (t-T)^{3IK-8K+3} dxdydz \\
S_{IKV}  & \approx  & \frac{\int \alpha_7 (t-T)^{3IK-8K+3} dxdydz}
{\int e^{2\nu} W (A-\beta_+) (t-T)^3 dxdydz}
\eea
where $\alpha_7$ is a positive function of $x,y$ and $z$.
Our results are depicted in
Tables (\ref{Indicators-Szekeres}) and
(\ref{DC-Szekeres}).
\\

\noindent {\em Elliptic class II Szekeres models:}
\\\\
The asymptotic evolution of $\rho$ and its time derivative are
in this case is given by 
\bea
\rho & \approx & \frac{6A}{\pi M^2 (\pi\beta_+-\beta_-)\psi_2} \\
\dot{\rho} & \approx & \frac{-6A}{\pi M^3
(\pi\beta_+-\beta_-)\psi_2^2}
\eea
where $\psi_2(z,t)=\left(\frac{t-T}{M}-2\pi\right)$.
Here we impose the restriction $\pi\beta_+-\beta_-<0$ to ensure 
the positivity of $\rho$ and 
by (\ref{Entropy-Szekeres}) we have
\bea
S_{IK}  & \approx & \int \alpha_8 |\psi_2|^{KI-\frac{10}{3}K+1} dxdydz\\
S_{IKV}  & \approx &  \frac{1}{27}\frac{\int \alpha_8 
|\psi_2|^{KI-\frac{10}{3}K+1} dxdydz}
{\int e^{2\nu} (\pi\beta_+-\beta_-) M \psi_2 dxdydz}
\eea
where $\alpha_8$ is a positive function independent of $t$.
The asymptotic behaviours of the indicators
for these models are identical to the corresponding models in class I
 (except in the parabolic case)
and the conditions for their
monotonic evolution
with time are
summarised in Tables (\ref{Indicators-Szekeres}) and (\ref{DC-Szekeres}).
Again the asymptotic behaviour of
$S_{IKV}$ cannot be deduced in general
for these models and therefore the corresponding entries in these Tables
were derived using pointwise
versions of the indicators.

\begin{table}[!htb]
\begin{center}
\begin{tabular}{c|l|ll}
\hline
\rule{0cm}{0.7cm}
&~~~~~~$Class$ II & \multicolumn{2}{c}{$Classes$ I \& II} \\
&&&\\
\hline
Indicators& $~~~~~~~~k=0$ & $~~~~~k=-1$ & $~~~k=+1$\\
\hline

$S_{IK}$  & \rule{0cm}{0.7cm} $(t-T)^{\frac{8}{3}IK-\frac{20}{3}K+2}$
 & $(t-T)^{3IK-8K+3}$ & $\psi_2^{IK-\frac{10}{3}K+1}$\\

$S_{IKV}$ &  \rule{0cm}{0.7cm} $(t-T)^{\frac{8}{3}IK-\frac{20}{3}K}$ & $(t-T)^{3IK-8K}$ & $\psi_2^{IK- \frac{10}{3} K}$\\

$B1$& \rule{0cm}{0.7cm} $(t-T)^{-\frac{4}{3}}$  & $(t-T)^{-2}$
 & $\psi_2^{-\frac{1}{3}}$\\

SW & \rule{0cm}{0.7cm} $(t-T)^{-\frac{2}{3}}$ & $const.$ 
& $\psi_2^{\frac{1}{3}}$\\

&&&\\
\hline
\end{tabular}
\caption[Indicators-Szekeres]{\label{Indicators-Szekeres}Asymptotic evolution of a number of
density contrast indicators  for the class I and class II 
Szekeres models. Columns 3 and 4 represent the
behaviours
for the hyperbolic and elliptic models
which are identical for both classes,
while the second column represents the behaviour for
class II models.}
\end{center}
\end{table}


\begin{table}[!htb]
\begin{center}
\begin{tabular}{l|cc|cc}
\hline
&&&&\\

&~~~~~~~~~~~$Class$ I & & \multicolumn{2}{c}{$Class$ II} \\

&&&&\\
\hline
 Models &  $S_{IK}$ & $S_{IKV}$ & $S_{IK}$ & $S_{IKV}$\\
\hline

$k=0$ & \rule{0cm}{0.7cm}$-$ & $-$ &
 $I> \frac{5}{2}-\frac{1}{K}$ &  $I> \frac{5}{2} $\\

$k=-1$ & \rule{0cm}{0.7cm} $I> \frac{8}{3}-\frac{1}{K}$  & $I>\frac{8}{3}$ &
 $I> \frac{8}{3}-\frac{1}{K}$ &  $I> \frac{8}{3} $ \\

$k=+1$ & \rule{0cm}{0.7cm} $I< \frac{10}{3}-\frac{1}{K}$ & $I<\frac{10}{3}$ &
 $I< \frac{10}{3}-\frac{1}{K}$ &  $I< \frac{10}{3}$\\
&&&&\\
\hline
\end{tabular}
\caption[DC-Szekeres]{\label{DC-Szekeres} Constraints on $I$ and $K$
in order to ensure $\dot{S}_{IK}>0;
\dot{S}_{IKV}>0$ asymptotically in 
class I and class II Szekeres models.}
\end{center}
\end{table}
\vskip .2in
To summarise, the results of this section indicate
that for both ever-expanding ($k=0$ and $k=-1$) and
recollapsing ($k=+1$) Szekeres models, 
we can always choose $I$ and $K$ such that the $S_{IK}$ and $S_{IKV}$
are asymptotically increasing both separately and
simultaneously. However, for some cases of interest,
such as e.g. $I=2$,
no such 
interval can be found  for which both sets of
indicators simultaneously have
this property.
Also as can be seen from Table (\ref{Indicators-Szekeres})
different indicators can, for different values of $I$ and
$K$, give different predictions
concerning the asymptotic homogenisation of these
models.
\\ 

We also note that the similarity between the results
of this section (for $k=+1$ and $k=-1$) and those for LT models 
(for $f>0$ and $f<0$) is partially due to the 
fact that the density functions $\rho$ 
has the same time dependence in both 
of these sub-families of LT and Szekeres models.
A nice way of seeing this, as was pointed out to us
by van Elst, is that for the
above dust models
the evolution equations for the density,
expansion, shear and 
electric Weyl curvature
constitute a closed dynamical system which
is identical for both models (see \cite{henk}).
This, however,
does not necessarily imply that
our indicators should also have 
identical time evolutions for all times
for both models, since they also include
$h^{ab}$ and $dV$ in their definitions.
It turns out that asymptotically
they are the same in the cases considered
here.
\section{Evolution of dimensionless indicators}
As they stand, indicators (\ref{comoving-ind}) are not
dimensionless in general. To make them so, we shall also
briefly consider their dimensionless analogues
given by
\be
\label{dimeless-ind}
S_{IKL}=\frac{S_{IK}}{V^L} 
\ee
where $L$ is a real number which depends on $I$ and
$K$ thus
\be
L=\frac{2}{3}IK-2K+1.
\ee
Asymptotic behaviour
of $S_{IKL}$
for LT and Szekeres models are summarised in Table (\ref{SIKL-Evolution}).
This demonstrates that there
are still intervals for $I$ and $K$ ($I \in \left]2,4\right[,
K\in\Re\setminus\{0\}$)
such that these
dimensionless indicators asymptotically increase with time.
In this way the results of the previous sections
remain qualitatively unchanged.

\begin{table}[!htb]
\begin{center}
\begin{tabular}{lccc}
 
\\
\hline
\\

Models  & Tolman & Szekeres I & Szekeres II  \\
\\
\hline
\\
Parabolic & -- & -- & $(t-T)^{\frac{4}{3}IK-\frac{8}{3}K}$   \\
\\
Hyperbolic &  $(t-a)^{IK-2K}$ & $(t-T)^{IK-2K}$
&   $(t-T)^{IK-2K}$ \\
\\
Elliptic &  $\phi_2^{\frac{1}{3}IK-\frac{4}{3}K}$
 & $\psi_2^{\frac{1}{3}IK-\frac{4}{3}K}$
& $\psi_2^{\frac{1}{3}IK-\frac{4}{3}K}$ \\
\\
\hline
\end{tabular}
\caption[SIKL-Evolution]{\label{SIKL-Evolution} Asymptotic behaviour
 of $S_{IKL}$
 for LT and Szekeres models. In the elliptic cases 
pointwise versions of the indicators were used.}
\end{center}
\end{table}
\section{Behaviour near the initial singularity}
So far we have studied the asymptotic behaviour of these models
at late times.
To further understand the possible monotonic 
behaviour of these indicators, 
it is of interest to study 
their behaviour near the initial singularities.
Our results are
summarised in
Table
(\ref{Initial-Sing}) and the constraints on $I$ and $K$
in order to ensure the non-divergence of $S_{IK}$, $S_{IKV}$ and 
$S_{IKL}$ near the singularities 
are given in Table
(\ref{Initial-Constraints}).

As can be seen from Table (\ref{Initial-Sing}), 
apart from the case of Szekeres II with decaying modes ($\beta_- \neq 0$),
indicators
have the same behaviour as we approach the origin
for all other
models. On the other hand, in
presence of decaying modes,
the constraints on $I$ and $K$ necessary to 
ensure initial non-divergence and final
monotonic increase are disjoint, ie, there
are no intervals of $I$ and $K$ such that these conditions 
are simultaneously satisfied.
This in turn seems to imply that there is an
incompatibility between the presence of 
decaying modes (see also \cite{Silk,Bonnor86,Goode-Wainwright})
and a unique density contrast arrow.

\begin{table}[!htb]
\begin{center}
\begin{tabular}{lllll}
\hline
\\
Indicators  & Tolman & Szekeres I & Szekeres II &  Szekeres II \\
& & & $(\beta_-\neq 0)$ & $(\beta_-=0)$\\
\\
\hline
\\
$S_{IK}$ &  $(t-a)^{2IK-4K+2}$ & $(t-T)^{2IK-4K+2}$
&  $(t-T)^{IK-\frac{10}{3}K+1}$ &  $(t-T)^{2IK-4 K+2}$ \\
\\
$S_{IKV}$ &  $(t-a)^{2IK-4K}$ & $(t-T)^{2IK-4K}$
&  $(t-T)^{IK-\frac{10}{3}K}$ &  $(t-T)^{2 IK-4K}$ \\
\\
$S_{IKL}$ &  $(t-a)^{\frac{2}{3}IK}$ & $(t-T)^{\frac{2}{3}IK}$
&  $(t-T)^{\frac{1}{3}IK-\frac{4}{3}K}$ &  $(t-T)^{\frac{2}{3}IK}$ \\
\\
\hline
\end{tabular}
\caption[Initial-Sing]{\label{Initial-Sing} The behaviour of 
the indicators $S_{IK}$, $S_{IKV}$ and $S_{IKL}$ in LT
and Szekeres models near the initial singularity.}
\end{center}
\end{table}
 

\begin{table}[!htb]
\begin{center}
\begin{tabular}{lccc}
\hline
\\
 Models & $S_{IK}$ & $S_{IKV}$ & $S_{IKL}$\\
\\
\hline
\\
 Lemaitre--Tolman & $I> 2-\frac{1}{K}$ &  $I> 2 $ & $IK>0$\\
\\
 Szekeres I &  $I> 2-\frac{1}{K}$ &  $I> 2 $ & $IK>0$\\
\\
 Szekeres II  ($\beta_-=0$) & $I> 2-\frac{1}{K}$ &  $I> 2 $ & $IK>0$ \\
\\
 Szekeres II ($\beta_-\ne 0$) & $I> \frac{10}{3}-\frac{1}{K}$ &  $I>\frac{10}{3}$
 & $I>4$\\
\\ 
\hline
\end{tabular}
\caption[Initial-Constraints]{\label{Initial-Constraints} Constraints on $I$ and $K$
in order to ensure the non-divergence and monotonic increase
of $S_{IK}$, $S_{IKV}$ and $S_{IKL}$ near the origin of time.}
\end{center}
\end{table}
\section{Some consequences}
In the following we shall briefly discuss some of the 
conceptual issues that the 
above considerations give rise to.
\subsection{Expanding versus recollapsing phases}
An interesting question regarding cosmological evolution is
whether the expanding and recollapsing evolutionary
phases do (or should) possess the same characteristics
in terms of density contrast indicators and in particular
whether or not the rise of structuration is expected
to increase monotonically throughout the history of the 
Universe independently of the cosmological
direction of its evolution, in other words, whether
the density contrast and the cosmological arrows
have the same sign.

The answer to this question is of potential importance
not only with regard to the issue of structure formation 
but also in connection with debates concerning
the nature of gravitational entropy and its likely behaviour
in expanding and contracting phases of the Universe 
(see e.g.
Hawking \cite{Hawking85}, Page \cite{Page} and also
references in \cite{Book-Arrow-Time}).
We note
that the underlying
intuitions in such debates seem to have come
mainly from considerations of the Friedman-Lemaitre-Robertson-Walker (FLRW)
models
together with some studies involving
small inhomogeneous perturbations about such models \cite{Hawkingetal93}. 
Our studies
here, though classical, can therefore provide potentially
useful complementary intuitions
by considering exact inhomogeneous models.
In particular, our results
in the previous sections raise a number
of points that can potentially 
inform this debate, amongst them:
\\

\begin{enumerate}

\item {\it Indicators in recollapsing models}: We have found that
different indicators behave differently in
expanding and recollapsing phases of cosmological evolution,
with some that remain monotonically increasing
in both phases and others which change their signs in the expanding
and recollapsing phases. 
In particular we find that there is a larger class of indicators
that remain monotonic in the ever-expanding models and in this
way we could say that
it is more difficult to maintain 
monotonic growth in 
recollapsing phases.

\item {\it Spatial dependence of turning time in recollapsing models}: 
As opposed to the closed FLRW
models, the turning time in inhomogeneous models 
can be position dependent. For example,
in the LT models
with $a=0$, the turning time is given by $t=\pi F/
(2f^\frac{3}{2})$, which in general depends on $r$.
In this sense there are epochs over which the Universe
possesses a multiplicity of cosmological arrows.

This raises the interesting question of 
whether
there can be observationally viable inhomogeneous cosmological
models which allow ranges of epochs and neighbourhoods,
such that over these epochs different
neighbourhoods (labelled
$\Sigma_N$ in the definition of the indicators (4), (6)
and (\ref{dimeless-ind}))
can give rise to 
different 
local density contrast arrows
within the same model.
\\

We note that in the 
ever-expanding models this problem is less likely
to arise,
which raises the question of whether this can
be taken as an argument in favour of
ever-expanding models which
more easily allow uni-directionality in their
density contrast arrows.
\end{enumerate}
\subsection{Connection to homogeneity}
\label{Connection-Homogeneity}
Another question is what is the connection
between homogeneity and the behaviour of density contrast indicators?
We start by recalling a result of Spero \& Szafron \cite{Spero-Szafron} 
according to which Szekeres models are spatially homogeneous
(with $S_{IK}=0 = S_{IKV}$) if the density
in these models is a function of $t$ only. 
Therefore in these models 
$S_{IK}=0 \Longrightarrow $ homogeneity.
Now in general
$S_{IK}=0$ may not imply homogeneity,
which raises the interesting question
of
what is the set of inhomogeneous models 
which satisfy this property? A related question has
been considered
by van Elst \cite{van-Elst} who points out some of the
difficulties involved.
We hope to return to this question in future.
\\

Another point in this regard is that
different indicators (even covariant ones)
can make contradictory statements 
about whether or not a model homogenises
asymptotically.
Here to illustrate this point 
we look at the following examples.
\\

First, we consider the parabolic LT model
studied by
Bonnor \cite{Bonnor74} in the form
\be
\label{metric}
ds^2=-dt^2+(t-a)^{4/3}
\left(\left(1+\frac{2\tilde{r}a'}{3(t+a)}\right)d\tilde{r}^2+\tilde{r}^2d\Omega^2
\right)
\ee
where
$a=a(\tilde{r})$.
Employing the indicator $B2$, 
Bonnor deduced that for fixed $\tilde{r}$
this model
approaches homogeneity, as $t\to \infty$, irrespective of its
initial conditions.
This indicator is not covariant. But even for 
covariant indicators (such as $S_{IK}$) one can
find 
ranges of $I$ and $K$ (namely $I >\frac{11}{3}-\frac{1}{K}$
and its complement) which give
opposite conclusions regarding asymptotic
homogenisation.
\\

As another example, we consider the hyperbolic LT models
studied by Bonnor \cite{Bonnor74} and given by $F=bf^{3/2}$, with
$b=$ constant. 
According to the $B1, B2, SL$ and $SW$ indicators
these models approach homogeneity, whereas there are
ranges of values of $I$ and $K$ (namely $I>\frac{10}{3}-\frac{1}{K}$)
for which
our measures $S_{IK}$ 
increase monotonically.
\\

We also take the hyperbolic class of
LT models with $a=0$, which were studied in
the subsection (\ref{LT-Hyperbolic}) above. As can be seen from
Table (\ref{DC-Tolman}), there are ranges of $I$ and $K$ 
for which the asymptotic behaviour
of both $S_{IK}$ and
$S_{IKV}$ can be the opposite to
the $B1$ indicator.
\\

Finally, It is also of interest to compare our results with the 
studies of asymptotic behaviour for Szekeres models.
An interesting result in this
connection is due to Goode \& Wainwright \cite{Goode-Wainwright},
according to which Szekeres models with $k=0,-1$ 
and $\beta_+ = 0$ become homogeneous asymptotically,
a result that was also obtained for the class II models 
by Bonnor \& Tomimura \cite{Bonnor-Tomimura}.
We recall that for
the models with simultaneous initial singularity considered
here, the assumption of $\beta_+ = 0$ reduces 
the class I models to that of
FLRW. For the class II models (with $\beta_-\ne 0$), the
asymptotic behaviour of our indicators depends on the choice of
the indices $I$ and $K$. In particular, to ensure the
asymptotic increase of the indicators $S_{IK}$, $S_{IKV}$, $S_{IKL}$
for $k=0$ and $k=-1$, it is necessary and sufficient to have 
$I> \frac{11}{3}-\frac{1}{K}$, $I>\frac{11}{3}$, $I>5$ and 
$I> \frac{10}{3}-\frac{1}{K}$, $I>\frac{10}{3}$, $I>4$
respectively.
\section{Conclusions}
An important assumption in gravitational physics concerns the
tendency of gravitational systems to become increasingly
lumpy with time. 
Here, we have tried to study this possibility
in the context of inhomogeneous
models of LT and Szekeres, by
using a number of
two parameter families of density contrast indicators. 
\begin{figure}
\centerline{\def\epsfsize#1#2{0.5#1}\epsffile{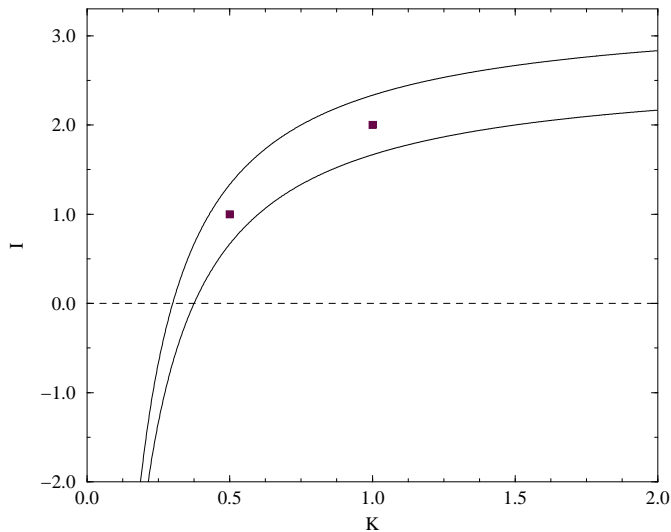}}
\caption{\label{Graph_dc}
The region between the curves gives the ranges of values of
$I$ and $K$ such that $S_{IK}$ is
increasing both near the singularity and asymptotically for all the models considered here.
The squares show the special examples of 
indicators with $I=1, K=1/2$ and $I=2, K=1$.}
\end{figure}

Given the arbitrary functions in these models, we have only been
able 
to establish conditions for monotonicity of our indicators
as we approach the origin and asymptotically.
Even though these are necessary but not sufficient conditions
for monotonicity for all times, nevertheless our studies 
illustrate some general points that seem to be 
supported by our all time study of a number of special models.
Our results show:
\begin{enumerate}
\item Different density contrast indicators can behave differently
even for the same model.
We find there is a larger class of indicators
that grow monotonically for ever-expanding models
than for the 
recollapsing ones. In particular,
in the absence of decaying modes ($\beta_-=0$),
we find that 
indicators exist
which grow monotonically with time
for all the models considered here.
Figure (1) gives a brief summary of our results
by depicting the
ranges of $I$ and $K$
such that $S_{IK}$ grow monotonically near the origin as
well as asymptotically 
for all the models considered here.
An example of a special indicator that
lies in this range is 
given by $K=1, I=2$ (a non pointwise
version of $B1$). However, the indicator 
given by
$K=1/2, I=2$ (which is
linear in the derivatives of density
\cite{Tavakol-Ellis}) does not
satisfy this property.  
\\

\item
If decaying modes exist
(i.e. $\beta_- \neq 0$), we find 
no such indicators which grow monotonically with time
for all the models considered here. 
Recalling a theorem of Goode and Wainwright \cite{Goode-Wainwright},
namely that $\beta_-=0$ is the necessary and sufficient condition
for the initial data (singularity) to be FLRW--like,
our results seem to imply that the presence
of monotonicity in the evolution of
density contrast indicators considered here
is directly related to the nature of 
initial data.
This is of potential relevance in the recent debates
regarding the nature of arrow of time (to the extent
that such an arrow can be identified with the density contrast arrow)
in ever expanding and recollapsing models
\cite{Hawking85,Page,Book-Arrow-Time}.

\item  Our considerations seem to indicate that the
notion of asymptotic
homogenisation as deduced from
density contrast indicators may not
be unique.
 
\item The possible spatial dependence of turning points in
inhomogeneous models can lead to multiplicity of
local density contrast arrows at certain epochs, which could
have consequences regarding
the corresponding cosmological arrow(s) of time.
\end{enumerate}

Finally we note that given the potential operational definability of
our indicators, it is interesting that the overall approach to homogeneity
may not necessarily be accompanied by a decrease
in the density contrast, as measured by the different 
indicators. In this way different covariant density contrast indicators
may give 
different insights as to what may be observed
asymptotically in such inhomogeneous models.
\vspace{.3in}
 
\centerline{\bf Acknowledgments}

\vspace{.3in}
We would like to thank Bill Bonnor, George Ellis, Henk van Elst
and Malcolm MacCallum for many helpful comments and discussions.
FCM wishes to thank Centro de Matem\'{a}tica da Universidade do Minho
for support and FCT (Portugal) for grant PRAXIS XXI BD/16012/98.
RT benefited from PPARC UK Grant No. L39094. 
\section{Appendix}
The different forms of the functions in Szekeres class I and II models
\cite{Goode-Wainwright}.

                              $$ Class~I$$
\be
\begin{array}{l}
R=R(z,t)\\
 f_{\pm}=f_{\pm}(t,z)\\
T=T(z)\\
 M=M(z)\\
e^\nu=f[a(x^2+y^2)+2bx+2cy+d]^{-1}\\
\end{array}
\ee
where functions in the metric are subjected to the conditions:
\be
\label{c1}
\begin{array}{ll}
ad-b^2-c^2=\frac{1}{4}\varepsilon & \varepsilon=0,\pm 1\\
A(x,y,z)=f{\nu_z}-k\beta_{+} \\
W(z)^2=(\varepsilon-kf^2)^{-1} \\
\beta_{+}(z)=-kfM_z(3M)^{-1} \\
\beta_{-}(z)=fT_z(6M)^{-1}
\end{array}
\ee
The functions $a, b, c, d$ and $f$ are all functions of $z$ that are
only
required to satisfy equations (\ref{c1}).

                                 $$Class~II$$
\be
\begin{array}{l}
R=R(t) \\
f_{\pm}=f_{\pm}(t)\\
 T=const.\\
 M=const.\\
e^\nu=[1+\frac{1}{4}k(x^2+y^2)]^{-1}\\
 W=1
\end{array}
\ee
\be
\begin{array}{lll}
A=\left\{\begin {array}{lll}
         e^\nu[a[1-\frac{1}{4}k(x^2+y^2)]+bx+cy]-k\beta_{+} & if & k=\pm
1\\
         a+bx+cy-\frac{1}{2}\beta_{+}(x^2+y^2) & if & k=0
         \end{array}
   \right.
\end{array}
\ee
in this case $a, b, c, \beta_{+}$ and $\beta_{-}$ are arbitrary
functions of $z$. The curvature is given by $k$.


\end{document}